\documentclass[prl,twocolumn,showpacs,superscriptaddress,preprintnumbers,amssymb]{revtex4}
\usepackage{graphicx}
\usepackage{dcolumn}
\usepackage{bm}
\usepackage{wasysym}

\def\vol{{\rm Vol}}

\newcommand{\beq}{\begin{equation}}
\newcommand{\eeq}{\end{equation}}
\newcommand{\beqn}{\begin{eqnarray}}
\newcommand{\eeqn}{\end{eqnarray}}

\begin{document}
\title{Strong-weak coupling self-duality in the two-dimensional quantum phase transition of $p+ip$ superconducting arrays}
\author{Cenke Xu}
\affiliation{Department of Physics, University of California,
Berkeley, CA 94720}
\author{J.~E.~Moore}
\affiliation{Department of Physics, University of California,
Berkeley, CA 94720}
\affiliation{Materials Sciences Division,
Lawrence Berkeley National Laboratory, Berkeley, CA 94720}
\date{\today}
\begin{abstract}
The 2D quantum phase transition that occurs in a square lattice of
Josephson-coupled $p \pm ip$ superconductors is an example of how four-body interactions in $d=2$ reproduce nonperturbative effects caused by two-body interactions in $d=1$.  The ordered phase has unconventional ``bond order'' of the
local $T$-breaking variable.  This problem can be analyzed
using an exact self-duality; this duality in classical notation is
the 3D generalization of the Kramers-Wannier duality of the 2D Ising model,
and there are similar exact dualities in dimensions $d \geq 3$.  We discuss the excitation spectrum and experimental signatures of the ordered and disordered phases,
and the relationship between our model and previously studied behavior of 2D boson models with four-boson interactions.
\end{abstract}
\maketitle

Many topical problems in condensed matter physics are described by
effective Hamiltonians with explicit three- or four-body
interaction terms, even though the underlying Coulomb interaction
between particles is only a two-body interaction.  An example is
the Pfaffian state of paired composite fermions in the fractional
quantum Hall effect~\cite{mooreread}, which is the exact ground
state of a three-body interaction~\cite{greiter}; this state has
been observed at $\nu=\frac{5}{2}$~\cite{willett}.  Regular
Josephson-junction arrays of a $p+ip$ superconductor like
Sr$_2$RuO$_4$ can be modeled by a multiple-spin lattice
Hamiltonian~\cite{moorelee}, as can several models of frustrated
magnetism~\cite{hermele} and superconductivity~\cite{paramekanti}.
This paper considers quantum effects on a classical four-spin
Hamiltonian as an improved model of a $p+ip$ superconducting array,
and presents several exact results on the resulting 2D quantum phase
transition.

This analysis is based on an exact strong-weak
coupling self-duality for multiple-spin interactions in high
dimensions, generalizing the Kramers-Wannier
duality~\cite{kramerswannier} of the classical Ising model in 2D
or the quantum Ising chain in 1D.  This higher-dimensional
self-duality continues recent developments~\cite{moessnersondhi,paramekanti} showing that phenomena that occur with two-body interactions in one quantum
dimension, like spin-charge separation, can also be realized by
three- or four-body interactions in two quantum dimensions.  Superconducting
arrays and frustrated magnets are important examples of this physics because they can generate three-, four-, or six-spin interactions without two-spin interactions, essentially because of unusual symmetries.  There are many known dualities in $d>2$ that
relate a strong-coupling regime of one model to a weak-coupling
regime of another model, e.g. the duality between the 3D Ising
model and 3D $\mathbb{Z}_2$ lattice gauge theory, but self-dualities
are quite rare in $d>2$~\cite{savit2}.

We start from the following classical model of Ising spins on a
square lattice: \beq \beta E = -K \sum_\square s^\square_1
s^\square_2 s^\square_3 s^\square_4 \label{clas2d} \eeq where
$s^\square_i = \pm 1$, $i=1,\ldots,4$ are Ising variables at the
four corners of one face of the lattice, and the sum is over all
faces.  This model was recently introduced~\cite{moorelee} to
understand the effects of frustrating geometric phases in a square
array of superconducting grains where each grain has either $p+ip$
or $p-ip$ order: the state of grain $i$ is described by both a
phase $\phi_i$ and an Ising variable $s_i = \pm 1$ that
determines the order parameter $p + i s_i p$.

The same geometric phases that led to the experimental determination of $d$
order in the cuprates~\cite{vanharlingen,tsuei} lead to
frustration of the superconductivity unless each plaquette of four
grains has an even number of Ising $+1$ spins (and an even
number of $-1$ spins)~\cite{moorelee}.  The phase $\Phi$ acquired by a Cooper pair moving around a plaquette is determined
by the states $s_i = \pm 1$ of the four grains at corners $i=1, \ldots, 4$: $\Phi = {\pi \over 2} (s_1 + s_2 + s_3 + s_4).$
This phase is equivalent to zero if the plaquette has an even number of $+1$ spins, and otherwise equivalent to $\pi$, which generates the local energy in (\ref{clas2d}) proportional to $s_1 s_2 s_3 s_4$.  Josephson weak links~\cite{davispackard} have been made in superfluid He$^3$, and
in one phase the symmetries of the weak link break the
symmetry group down to $p\pm ip$.

The classical model (\ref{clas2d}) also describes the
two-dimensional ``right-angle water'' ice model~\cite{ziman} and
maps onto to a case of the eight-vertex
model~\cite{baxterbook,moorelee}.  Note that (\ref{clas2d}) is not
ordinary 2D $\mathbb{Z}_2$ gauge theory, where the sum is over
bond variables $\sigma_i$ around each face.  The
overall symmetry group is much smaller for
(\ref{clas2d}): $\vol(G^\prime) = 2^{N_x + N_y}$ rather than
the full gauge group $\vol(G) = 2^{N_x N_y}$ of $\mathbb{Z}_2$
gauge theory.  The problem (\ref{clas2d}) has a one-dimensional
ground-state degeneracy $2^{N_x+N_y}$ even with no physical boundary (e.g. on a
torus).  The model can be solved in the thermodynamic limit for
all $K$: its free energy per face is just $\beta f=-\log(2 \cosh K)$,
since all the face variables can be chosen independently. 
The model is equivalent to the 1D Ising model and has no
phase transition.

The model becomes quantum-mechanical in the presence of a transverse
magnetic field: \beq H = -K \sum_\square \sigma^z_1 \sigma^z_2
\sigma^z_3 \sigma^z_4 - h \sum_i \sigma_i^x. \label{quant2d} \eeq
Here again the $K$ interaction is around a plaquette but now the
spin is a quantum spin-half and the $\sigma$ are Pauli matrices.
In the superconducting array realization, the magnetic field $h$
corresponds to tunneling between the two order parameters $p\pm
ip$, which will in the limit of strong tunneling induce a single real order parameter $p_x$.  Application of pressure in Sr$_2$RuO$_4$ is found experimentally to drive the system toward a real $p$ state~\cite{mackenziemaeno}, but the explanation of this effect is unclear.  The model (\ref{quant2d}) is clearly one of the simplest possible 2D lattice quantum Hamiltonians with four-spin interactions.  The main weakness of the model (\ref{quant2d}) is that in the real system at low $T$ there is a long-ranged
vortex-vortex interaction between frustrated
plaquettes~\cite{moorelee}; the Hamiltonian (\ref{quant2d})
corresponds to treating the core energy of a vortex but not its
interaction with other vortices, as might be appropriate at higher
temperatures.

We now show that the quantum model (\ref{quant2d}) has a
phase transition at zero temperature when $K/h$ is
exactly one.  For simplicity we will give the model's self-duality here in its classical form;  the same duality can be shown directly in the quantum model (\ref{quant2d}) and interchanges $K$ and $h$~\cite{cenkelong}.  The 3D anisotropic classical model
\beq \beta E = -{\tilde K} \sum_\square s^\square_1 s^\square_2
s^\square_3 s^\square_4 - J_z \sum_b s^b_1 s^b_2, \label{clas3d}
\eeq where $\square$ ranges over all
plaquettes in the $xy$ planes, and $b$ ranges over all bonds in the $z$ direction, will be shown to have a phase transition along the line \beq \sinh 2 {\tilde K}
\sinh 2 J_z = 1. \label{transloc} \eeq
The connection between coupling constants in the
classical model and in the quantum model is
standard~\cite{sachdev}: \beq {\tilde K} = a K,\quad e^{-2 J_z} =
\tanh(a h), \quad T = {1 \over M a} \eeq where $a$ is the lattice spacing
and $M$ the number of sites in the $z$ direction of the classical
model, and $T$ is the temperature in the quantum case.

The quantum-classical mapping becomes exact in the limits $T \rightarrow 0$, $a
\rightarrow 0$, and $M \rightarrow \infty$.  Knowing the phase
transition line in the classical model (\ref{transloc}) fixes
the quantum transition because in the above limits \beq
\sinh 2 {\tilde K} \sinh 2 J_z \rightarrow {\tilde K} e^{2 J_z} =
K / h = 1. \eeq

\begin{figure}
\includegraphics[width=2.5in]{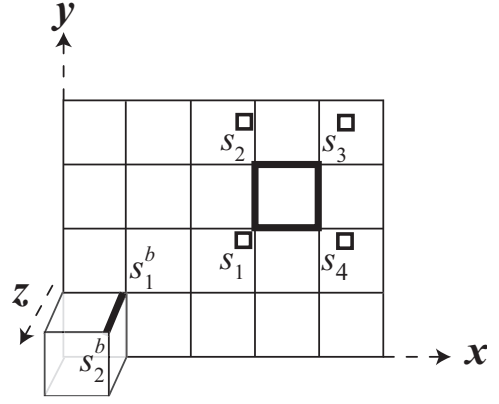}
\caption{The classical anisotropic 3D problem that describes the
quantum critical point of the model (\ref{quant2d}).  The two
types of interactions (shaded bonds) are plaquette interactions in
the planes normal to ${\bf \hat z}$, and bond interactions along
${\bf \hat z}$. } \label{figone}
\end{figure}

The partition function of the above model is \beqn Z =
\sum_{\{s\}} e^{-\beta E} &=& \sum_{\{s\}} \prod_\square
\Big[(\cosh {\tilde K} + s^\square_1 s^\square_2 s^\square_3
s^\square_4 \sinh {\tilde K})\cr &&\times \prod_b (\cosh J_z +
s^b_1 s^b_2 \sinh J_z)\Big]. \eeqn Introduce face variables
$k_\square = 0,1$ and bond variables $k_b = 0, 1$, and define $c_0
= \cosh {\tilde K}$, $c_1 = \sinh {\tilde K}$, $d_0 = \cosh J_z$,
$d_1 = \sinh J_z$.  Then
\beqn Z &=& \sum_{\{s\}} \sum_{k_\square}
\sum_{k_b}  \Big[ \left( \prod_{\square}  c_{k_\square}
(s^\square_1 s^\square_2 s^\square_3 s^\square_4)^{k_\square}
\right) \cr && \times \left( \prod_b d_{k_b} (s^b_1
s^b_2)^{k_b}\right)\Big]. \eeqn
Now the spin sum can be evaluated:
for each spin the result is 2 if the spin is raised to an even
power, and 0 otherwise.  $Z$ is a constrained sum over the $k$
variables: \beq Z = 2^N {\sum_{k_\square,k_b}}^\prime \left(
\prod_{\square} c_{k_\square} \right) \left(\prod_b d_{k_b}
\right). \eeq Here $N=N_x N_y N_z$ is the total number of sites.
Each site of the original lattice appears via 4
face terms and 2 bond terms.  The constraint is that the sum of
the six $k$ variables be an even number for every site.

Now introduce dual variables to solve the constraint.  The dual
spins $\sigma$ are located at the centers of the fundamental cubes
of the original cubic lattice.  For a site $i$ of the original
lattice, its four neighboring spacelike faces are pierced by four
vertical bonds of the dual lattice, and for each piercing bond $b$
of the dual lattice fix the relation $k_\square = \frac{1}{2}(1 -
\sigma^b_1 \sigma^b_2)$.  Each of the two vertical bonds $b$
containing site $i$ pierces a spacelike face $\square$ of the dual
lattice, and we set $k_b = \frac{1}{2}(1 - \sigma^\square_1
\sigma^\square_2 \sigma^\square_3 \sigma^\square_4).$

These variables satisfy the constraint since the eight
dual lattice sites $\sigma_1,\ldots,\sigma_8$ around an original site satisfy
\beqn && k_{\square 1} + k_{\square 2}
+ k_{\square 3} + k_{\square 4} + k_{b 1} + k_{b 2} = 3 -
\frac{1}{2}(\sigma_1 \sigma_2 \sigma_3 \sigma_4 \cr&&+ \sigma_5
\sigma_6 \sigma_7 \sigma_8 + \sigma_1 \sigma_5 + \sigma_2 \sigma_6
+ \sigma_3 \sigma_7 + \sigma_4 \sigma_8) \equiv
0\,\rm{mod}\,2.\cr&& \eeqn This holds if all spins are up, and
flipping any spin changes the sum by an even number.  Next we need
to find how many dual spin configurations correspond to one
configuration of the $k$ variables.  The answer is just the size
of the gauge group $\vol(G^\prime) = 2^{N_x + N_y}$, since once
the dual spin configuration is set on a spacelike plane, the
vertical bonds fix the configuration everywhere else.

The last step is to calculate the dual couplings.
Writing \beq c_k = k \sinh {\tilde K} + (1-k) \cosh {\tilde K},
\eeq for a face $\square$ of the original lattice, pierced by
Ising bond $b$ in the dual problem, \beqn c_{k_\square} =& {1 +
\sigma^b_1 \sigma^b_2 \over 2} \cosh {\tilde K} + {1 - \sigma^b_1
\sigma^b_2 \over 2} \sinh {\tilde K} \cr =& {e^{{\tilde K}} \over
2} (1 + \sigma^b_1 \sigma^b_2 \tanh J_z^*) = {1 \over \sqrt{2
\sinh 2J_z^*}} e^{J_z^* \sigma^b_1 \sigma^b_2} \eeqn where $\tanh
J_z^* = e^{-2 {\tilde K}}$.  By the same process \beq d_{k_b} = {1
\over \sqrt{2 \sinh 2 {\tilde K}^{*}}} e^{{\tilde K}^{*}
\sigma^\square_1\sigma^\square_2 \sigma^\square_3
\sigma^\square_4} \eeq with $\tanh {\tilde K}^{*} = e^{-2 J_z}$.

Now we combine the above results: the number of bonds
parallel to ${\bf \hat z}$ and the number of spacelike faces are
both $N = N_x N_y N_z$, so we have up to boundary terms
\beq Z({\tilde K},J_z) =
{Z({\tilde K}^*,J_z^*) \over \vol(G^\prime) \sinh(2 J_z^*)^{N/2} \sinh(2
{\tilde K}^*)^{N/2}}. \eeq More precisely, we have shown that in the limit
$N \rightarrow \infty$, \beqn {\log Z(\tilde K,J_z) \over N} &=&
{\log Z({\tilde K}^*,J_z^*) \over N} - \frac{1}{2} \log \sinh(2 {\tilde K}^*) \cr&&
- \frac{1}{2} \log \sinh(2 J_z^*). \eeqn The duality relation for
the couplings is \beq \sinh (2 J_z^*) = {1 \over \sinh(2 {\tilde
K})},\quad \sinh(2 {\tilde K}^*) = {1 \over \sinh(2 J_z)}. \eeq This
determines the entire phase boundary line in the $({\tilde K},
J_z)$ plane.  It also shows that there is a hidden symmetry
between the couplings ${\tilde K}$ and $J_z$.  The
same construction leads to an exact duality of the 4D model with
eight-spin interactions around spacelike cubes, plus a bond
interaction in the fourth dimension; and similarly for any $d \geq
2$.  The self-dual point is ${\tilde K} = J_z = K_c = \frac{1}{2}
\log(1 + \sqrt{2}).$ If there is only one phase transition in the
model at finite coupling, it must be at
the self-dual line.  The self-duality of this model is similar to
the classical anisotropic self-dual Villain $\mathbb{Z}_N$ models
studied in~\cite{savit2}.

We remark that the model can be solved if the system has only one row of
spacelike plaquettes: the bond variables $b_x = s_{x,1} s_{x,2}$
become spins in an anisotropic 2D Ising model, and 
with $e^{2 J^\prime} = \cosh 2 J_z$,
\beq {\log
Z_{N_x \times 2 \times N_z}({\tilde K},J_z) \over N_x N_z} = {\log
Z^{2DI}_{N_x \times N_z} (2 {\tilde K},J^\prime) \over N_x N_z}.
\eeq


If there is a single second-order transition in the 3D classical model, then there is a second-order transition at $K=h$ in the quantum model.  The spontaneous order that develops across the transition is unconventional ``bond order''.  At $h=0$, the ground states all have long-range order along every row of the
vertical bond $\sigma^z_i \sigma^z_{i+{\hat y}}$, and of horizontal bonds along each column.  Another way to describe the $h=0$ ordered state is using the Wilson operator for the
product of spins around a loop ${\cal L}$, \beq W({\cal L}) =
\prod_{i \in {\cal L}} \sigma^z_i. \eeq In any ground state,
$\langle W({\cal L}) \rangle = 1$.  We will use bond order below
to make a connection between the order parameter and
the excitation spectrum.  The two orders are connected
since a closed loop contains an even number
of bonds from each row and column.


A Peierls-type argument can be used to show that there is an ordered
phase of the classical 3D model at low temperature, and
hence at least one phase transition.  We have performed
Monte Carlo and high-temperature series calculations to
check whether the above model has a single second-order
phase transition (which must then lie at the self-dual point $K_c$).  The results are consistent with this picture, but the single-spin Monte Carlo algorithm
becomes very slow close to criticality, as in~\cite{chamonunpub}.

The high-temperature series proves that this model
does not have the same free energy per site as an Ising model, as
might have been suspected since the classical 2D model with face
interactions (\ref{clas2d}) has the same free energy per site as the 1D two-spin Ising
model. The first terms for the classical symmetric model
(${\tilde K} = J_z$) are
\beq c = T {\partial s \over \partial T}
= -T{\partial^2 F \over N (\partial T)^2} = 2 {\tilde K}^2 - 2
{\tilde K}^4 + 94 {\tilde K}^6/3 + O({\tilde K}^8). \eeq
This
differs from the $d \geq 2$ two-spin Ising model at order ${\tilde
K}^4$ and the $d=1$ two-spin Ising model at order ${\tilde K}^6$.


The quantum model (\ref{quant2d}) has an infinite but non-extensive set of conservation
laws: along any one of the $N_y$ rows, say ${\cal R}$, the product
\beq
{\hat O}_{\cal R} = \prod_{i \in {\cal R}} \sigma^i_x
\label{conslaw}
\eeq
commutes with the Hamiltonian (\ref{quant2d}), and similarly for each of $N_x$ columns.  These $N_x+N_y$ conserved quantities are related to the ground-state degeneracy in the ordered state.  There are $2^{N_x+N_y}$ sectors of the theory, labeled by the eigenvalues of the operators (\ref{conslaw}).  In the large-$h$ phase, there is
a single ground state invariant under $N_x+N_y$ transformations that each act on all the spins in one row or column via
\beq
\sigma_x \rightarrow \sigma_x, \quad \sigma_y \rightarrow -\sigma_y, \quad \sigma_z \rightarrow -\sigma_z.
\eeq

At the transition $K=h$ in the thermodynamic limit, these symmetries are spontaneously broken and there are $2^{N_x+N_y}$ degenerate ground states once $K>h$.  In a finite system, there is no spontaneous symmetry breaking and ground-state degeneracy on the ordered side $K>h$ will be split by an amount exponentially small in $\min(N_x,N_y)$.  The breaking of these many symmetries at a single transition occurs because of the infinite number of conservation laws.  In the language of hard-core bosons, the charge
along each row or column is conserved modulo 2, just as in the
Bose metal model discussed in~\cite{paramekanti}.

Now we consider the excitation spectrum in the two phases of the
quantum model.  Recall the familiar quantum Ising
chain ($d=1$)~\cite{sachdev}: the lowest excitation in the
ordered phase with periodic boundary conditions is to flip one
spin from the ground state, so the first excited state contains
two bad bonds. These two bad bonds can be separated into two
kinks, with a string of flipped spins between them.  In the large
$h$ limit, all the spins point along ${\bf \hat x}$ and the
one-particle state is a flipped spin with momentum
and kinetic energy, with \beq \epsilon_k = K h\left[2-2/h\cos(k)+O(1/h^2)\right].
\label{dispersion}\eeq For our model in 2D, in the large $K$ limit
the system stays near the ground state manifold, and the first
excited state locally connected to the ground state
is obtained by flipping one spin. This flipped spin
results in four bad plaquettes (or four $\mathbb{Z}_2$
vortices).  As in the $d=1$ case, this excitation can
disintegrate into four fractional excitations, which become four
vertices of a rectangle with all the interior spins flipped.

The large-$h$ limit requires more attention.  In the ${\bf \hat x}$ basis, it is clear that the
single flipped spin is nondispersive because of the ${\mathbb
Z}_2$ conservation laws.  The single flipped spin in the large-$h$ limit is equivalent via duality to a single bad plaquette in the large-$K$ limit.  Instead, the lowest mobile state is a
flipped bond: the flipped bond can hop in one direction, the
direction perpendicular to the bond, and its dispersion relation
is the same as that in the $d=1$ case (\ref{dispersion}).  These
one-particle states can scatter off each other, and because they
hop unidirectionally, this kind of scattering is very similar to
1D scattering.  The ordered state at small $h$ is a condensation of bonds.

Direct experimental observation of the order-parameter state of a superconducting
grain is currently only possible for isolated superconducting grains.  However, there are several experimental signatures of the bond ordering predicted above, even though in the ground state the system has no frustrating fluxes and hence is uniform from the superconducting point of view.  The gapless fluctuations near the transition will modify the specific heat and transport in the system: transport will be attenuated near the critical point by scattering off the fluctuations, while specific heat will show a peak.  The true spin action in a $p+ip$ superconducting array is more complicated and includes long-ranged interactions between frustrated plaquettes~\cite{moorelee}, but it has the same gauge symmetries as (\ref{quant2d}).  If grain orders could be measured at the array boundary, the directionality of the predicted bond ordering would be seen, and the Ising variables of each neighboring pair of grains develop a symmetry-breaking correlation at the transition.

An unresolved question is whether (\ref{quant2d}) is
a free-fermion model like 1D quantum Ising.  We have not found additional conservation laws beyond ordinary symmetries and the infinitely many $\mathbb{Z}_2$ laws
described above.  A solution of this 2D quantum model would give further insight into the physics of explicit four-point interactions.

The authors acknowledge helpful conversations with D.-H. Lee.  J.
E. M. was supported by NSF DMR-0238760 and the Hellman Foundation.
This research used NERSC resources (DOE contract
DE-AC03-76SF00098).

\bibliographystyle{../Newliou/apsrev}
\bibliography{../bigbib}

\end{document}